\documentstyle[12pt,epsfig,cite,amsmath,amsfonts,amssymb]{article}
\newcounter{subfig}

\newcommand{\Prob}{\textrm{Pr}}
\newcommand{\sP}{\textrm{Pr}}
\newcommand{\tE}{\textrm{E}}
\newcommand{\ov}{\overline}

\newcommand{\xm}{{\mathbf x}_m}
\newcommand{\ym}{{\mathbf y}}
\newcommand{\pv}{{\mathbf p}}
\newcommand{\ud}{{\mathrm d}}
\newcommand{\err}{\varepsilon}
\newcommand{\beq}{\begin{equation}}
\newcommand{\enq}{\end{equation}}
\newcommand{\beqa}{\begin{eqnarray}}
\newcommand{\enqa}{\end{eqnarray}}
\newcommand{\beqn}{\begin{eqnarray*}}
\newcommand{\enqn}{\end{eqnarray*}}
\newcommand{\no}{\nonumber}

\newtheorem{theorem}{Theorem}
\newtheorem{lemma}[theorem]{Lemma}
\newtheorem{definition}[theorem]{Definition}
\newenvironment{proof}{{\sl Proof\/}:\ \ }{\qed\vspace{\baselineskip}}
\newcommand{\qed}{\hfill $\Box$}

\begin{document}

\title{On the Error Exponents of ARQ Channels with Deadlines}
\author{Praveen Kumar Gopala, Young-Han Nam and Hesham El Gamal\footnote{The
authors are with the ECE Dept., The Ohio State University, Emails: \{gopalap,namy,helgamal\}@ece.osu.edu.}}
\maketitle

\begin{abstract}
We consider communication over Automatic Repeat reQuest (ARQ)
memoryless channels with deadlines. In particular, an upper bound
$L$ is imposed on the maximum number of ARQ transmission rounds.
In this setup, it is shown that incremental redundancy ARQ
outperforms Forney's memoryless decoding in terms of the
achievable error exponents.
\end{abstract}

\section{Introduction}
In \cite{Burn}, Burnashev characterized the maximum error exponent
achievable over discrete memoryless channels (DMCs) in the
presence of perfect output feedback. Interestingly, Forney has
shown that even one-bit feedback increases the error exponent
significantly~\cite{Forn}. More specifically, Forney proposed a
memoryless decoding scheme, based on the erasure decoding principle,
which achieves a significantly higher error exponent than that
achievable through maximum likelihood (ML) decoding without
feedback \cite{Gall}. In Forney's scheme, the transmitter sends
codewords of block length $N$. After receiving each block of $N$
symbols, the receiver uses a reliability-based erasure decoder and
feeds back one ACK/NACK bit indicating whether it has
accepted/erased the received block, respectively. If the
transmitter receives a NACK message, it then re-transmits the same
$N$-symbol codeword. After each transmission round, the receiver
attempts to decode the message using {\bf only} the latest $N$
received symbols, and discards the symbols received previously.
This process is repeated until the receiver decides to accept the
latest received block and transmits an ACK message back to the
transmitter.

It is intuitive to expect a better performance from schemes that
do not allow for discarding the previous observations at the
decoder, as compared with memoryless decoding. Our work here is
concerned with one variant of such schemes, i.e., Incremental
Redundancy Automatic Repeat reQuest(IR-ARQ)~\cite{Caire}. We
further impose a deadline constraint in the form of an upper bound
$L$ on the maximum number of ARQ rounds. In the asymptotic case
$L\rightarrow \infty$, we argue that IR-ARQ achieves the same error
exponent as memoryless decoding, denoted by $E_F(R)$. On the other
hand, for finite values of $L$, it is shown that IR-ARQ generally
outperforms memoryless decoding, in terms of the achievable error
exponents (especially at high rates and/or small values of $L$).
For example, we show that $L=4$ is enough for IR-ARQ to achieve
$E_F(R)$ for any binary symmetric channel (BSC), whereas the
performance of memoryless decoding falls significantly short from
this limit.

The rest of this correspondence is organized as follows. In
Section~\ref{infL}, we briefly review the memoryless decoding
scheme without any delay constraints, and argue that allowing for
memory in decoding does not improve the error exponent. The
performance of the memoryless decoder and the incremental
redundancy scheme, under the deadline constraint, is characterized
in Section~\ref{finL}. In Section~\ref{examples}, we consider
specific channels (like the BSC, VNC and AWGN channels) and
quantify the performance improvement achieved by incremental
redundancy transmission. Finally, some concluding remarks are
offered in Section~\ref{conc}.

\section{The ARQ Channel} \label{infL}
We first give a brief overview of the memoryless decoding scheme
proposed by Forney in \cite{Forn}. The transmitter sends a
codeword $\xm$ of length $N$, where $m \in \{ 1, \cdots, M\}$.
Here $M$ represents the total number of messages at the
transmitter, each of which is assumed to be equally likely. The
transmitted codeword reaches the receiver after passing through a
memoryless channel with transition probability $p(y|x)$. We denote
the received sequence as $\ym$. The receiver uses an erasure
decoder which decides that the transmitted codeword was $\xm$ iff
$\ym \in {\mathcal R}_m$, where \beq \label{Rm} {\mathcal R}_m ~=~
\left\{ \ym ~:~ \frac{p(\ym|\xm)}{\sum_{k \ne m} p(\ym|{\mathbf
x}_k)} ~\ge~ e^{NT} \right\}, \enq where $T \ge 0$ is a
controllable threshold parameter. If (\ref{Rm}) is not satisfied
for any $m \in \{1,\cdots,M\}$, then the receiver declares an
erasure and sends a NACK bit back to the transmitter. On
receiving a NACK bit, the transmitter repeats the codeword
corresponding to the same information message. We call such a
retransmission as an ARQ round. The decoder discards the earlier
received sequence and uses only the latest received sequence of
$N$ symbols for decoding (memoryless decoding). It again applies
the condition (\ref{Rm}) on the newly received sequence and again
asks for a retransmission in the case of an erasure. When the
decoder does not declare an erasure, the receiver transmits an ACK
bit back to the transmitter, and the transmitter starts sending
the next message. It is evident that this scheme allows for an
infinite number of ARQ rounds. This scheme can also be implemented
using only one bit of feedback (per codeword) by asking the
receiver to only send back ACK bits, and asking the transmitter to
keep repeating continuously until it receives an ACK bit. Since
the number of needed ARQ rounds for the transmission of a
particular message is a random variable, we define the error
exponent of this scheme as follows.
\begin{definition}
The error exponent $E(R)$ of a variable-length coding scheme is
defined as \beq E(R) ~=~ \limsup_{N \to \infty} ~- \frac{\log
~{\em \sP(E)}} {\ov{\tau}} ~, \enq where ${\em \sP(E)}$ denotes
the average probability of error, $R$ denotes the average
transmission rate, and $\ov{\tau} = (\ln M/R)$ is the average
decoding delay of the scheme, when codewords of block length $N$
are used
in each ARQ transmission round. \\[0.001in]
\end{definition}
The probability of error of the decoder in (\ref{Rm}), after each
ARQ round, is given by \cite{Forn}
\[ \sP(\err) ~=~ \sum_{m} \sum_{k \ne m} \sum_{\ym \in R_k} p(\ym,\xm) ~, \]
and the probability of erasure is given by
\[ \sP(X) ~=~ \left(\sum_m \sum_{\ym \notin R_m} p(\ym,\xm) \right) ~-~
\sP(\err) ~. \] It is shown in \cite{Forn} that these
probabilities satisfy \beq \label{upper} \sP(X) \le e^{-N
E_1(R_1,T)} \quad \textrm{and} \quad \sP(\err) \le e^{-N
E_2(R_1,T)} ~, \enq where $R_1 = (\ln M / N)$ denotes the rate of
the first transmission round,
\beq E_2(R_1,T) ~=~ E_1(R_1,T) + T, \enq
and $E_1(R_1,T)$ is given at high rates by \cite{Forn}
\beqa 
E_1(R_1,T) &=& \max_{0 \le s\le \rho \le 1, \pv} ~ E_o(s,\rho,\pv)
- \rho R_1 - sT , \label{E1eqh} \\ E_o(s,\rho,\pv) &=& - \log \int \left(
\int p(x) p(y|x)^{(1-s)} \ud x \right) \left( \int p(x) p(y|x)^{
(s/\rho)} \ud x \right)^{\rho} ~\ud y ~, \label{E0h} \enqa
and at low rates by
\beqa
E_1(R_1,T) &=& \max_{0 \le s \le 1,\rho \ge 1, \pv} ~ E_x(s,\rho,\pv)
- \rho R_1 - sT , \label{E1eql} \\ E_x(s,\rho,\pv) &=& - \rho ~\log \int \int 
p(x) p(x_1) \left( \int p(y|x)^{(1-s)} p(y|x_1)^s \ud y \right)^{(1/\rho)}
~\ud x ~\ud x_1 ~, \label{Exl} \enqa
where $\pv = \{ p(x), \forall x\}$ denotes the input probability distribution
(We note that for discrete memoryless channels, the integrals in (\ref{E0h})
and (\ref{Exl}) are replaced by summations). The average decoding delay
$\ov{\tau}$ of the memoryless decoding scheme is given by \beqn
\ov{\tau} & = & \sum_{k=1}^{\infty} kN ~
\Prob(\textrm{Transmission stops after $k$ ARQ rounds}) \\ & = &
\sum_{k=1}^{\infty} kN ~ [\sP(X)]^{(k-1)}
 [1-\sP(X)] ~~ = ~~  \frac{N}{1 - \sP(X)} ~, \enqn
which implies that the average effective transmission rate is
given by \beq R ~ = ~ \frac{\ln M}{\ov{\tau}} ~ = ~ \left(
\frac{\ln M}{N} \right) [1 - \sP(X)] ~ = ~ R_1 [ 1 -
\sP(X)]~.\label{rate-eq} \enq It is clear from (\ref{upper}) and
(\ref{rate-eq}) that $R \to R_1$ as $N \to \infty$ if $E_1(R_1,T)
> 0$. The overall average probability of error can be now computed
as \beq \label{maineq} \sP(\tE) ~~ = ~ \sum_{k=1}^{\infty}
[\sP(X)]^{(k-1)} ~ \sP(\err) ~~=~~  \sP(\err) \left[ 1 + o(1)
\right], \enq where the second equality follows from (\ref{upper})
when $E_1(R_1,T)>0$. It is, therefore, clear that the error
exponent achieved by the memoryless decoding scheme is
\[ E(R) ~=~  \limsup_{N \to \infty} ~- \frac{\log \left(\sP(\err)[1+o(1)]\right)}
{\ov{\tau}} ~ \ge ~ E_2(R,T). \] It is shown in \cite{Forn} that
choosing the threshold $T$ such that $E_1(R_1,T) \to 0$ maximizes
the exponent $E_2(R_1,T)$ while ensuring that $R \to R_1$ as $N
\to \infty$. This establishes the fact that the memoryless decoding
scheme achieves the feedback error exponent $E_F(R)$ defined as
\beq E_F(R) ~ \triangleq ~ \lim_{E_1(R,T) \to 0} ~E_2(R,T) ~~=~
\lim_{E_1(R,T) \to 0} ~T ~. \enq

At this point, it is interesting to investigate whether a better
error exponent can be achieved by employing more complex receivers
which exploit observations from previous ARQ rounds in decoding
(instead of discarding such observations as in memoryless decoding).
Unfortunately, it is easy to see that this additional complexity
does not yield a better exponent in the original setup considered
by Forney~\cite{Forn}. The reason is that, as shown in
(\ref{maineq}), the overall probability of error in this setup is
dominated by the probability of error $\sP(\err)$ in the first
transmission round. So, while our more complex decoding rule might
improve the probability of error after subsequent rounds, this
improvement does not translate into a better error exponent. In
the following section, however, we show that in scenarios where a
strict deadline is imposed on the maximum number of feedback
rounds, significant gains in the error exponent can be reaped by
properly exploiting the received observations from previous ARQ
rounds (along with the appropriate encoding strategy).

\section{ARQ with a Deadline} \label{finL}
In many practical systems, it is customary to impose an upper
bound $L$ on the maximum number of ARQ rounds (in our notation, $L
\ge 2$ since we include the first round of transmission in the
count). Such a constraint can be interpreted as a constraint on
the maximum allowed decoding delay or a deadline constraint. With
this constraint, it is obvious that the decoder can no longer use
the rule in (\ref{Rm}) during the $L^{th}$ ARQ round. Therefore,
after the $L^{th}$ round, the decoder employs the maximum
likelihood (ML) decoding rule to decide on the transmitted codeword.
We denote the probability of error of the ML decoder by
$\sP^{(ML)}(\err)$.
\subsection{Memoryless Decoding} \label{mdL}
The following theorem characterizes lower and upper bounds on the
error exponent achieved by the memoryless decoding scheme, under
the deadline constraint $L$.

\begin{theorem} \label{md}
The error exponent $E_{MD}(R,L)$ achieved by memoryless decoding,
under a deadline constraint $L$, satisfies\footnote{We note that a
tighter lower bound may be obtained by using the expurgated
exponent $E_{ex}(R)$ instead of the random coding exponent
$E_r(R)$ at low rates. This observation will be used when
generating numerical results.} (for $0 \le R \le C$) \beq
\label{errexp} E_r(R) + (L-1) \left[ \max_{0 \le s \le \rho \le 1,
\pv} \left( \frac{ E_o(s,\rho,\pv) - \rho R -s E_r(R)}{1+s(L-2)}
\right) \right] ~ \le ~ E_{MD}(R,L)  ~ \le ~ L E_{sp}(R) ~, \enq
where $E_r(R)$ and $E_{sp}(R)$ denote the random coding and sphere
packing exponents of the memoryless channel, and $E_o(s,\rho,\pv)$
is as given in (\ref{E0h}).
\end{theorem}
\begin{proof}
The average decoding delay of memoryless decoding is given by
\beqa \ov{\tau} & = & \left( \sum_{k=1}^{L-1} kN ~
[\sP(X)]^{(k-1)} [1-\sP(X)] \right) + LN [\sP(X)]^{(L-1)} \no \\ &
= & \left( \sum_{k=0}^{L-1} (k+1)N
[\sP(X)]^k \right) - \left( \sum_{k=1}^{L-1} kN [\sP(X)]^k \right) \no \\
& = & N \left( \sum_{k=0}^{L-1} ~ [\sP(X)]^{k} \right) ~ = ~ N \left[ 1 +
o(1) \right]~, \enqa
where the last equality follows from (\ref{upper}) when $E_1(R_1,T)>0$. Thus
the average effective transmission rate is given by
\[ R ~=~ \frac{\ln M}{\ov{\tau}} ~=~ \frac{\ln M}{N [1 + o(1)]} ~\to~ R_1, \]
as $N \to \infty$ when $E_1(R_1,T) > 0$. The average probability of error is given by
\beqa \sP_{MD}(\tE) & = & \sum_{k=1}^{L-1} [\sP(X)]^{(k-1)} ~
\sP(\err) ~+~ [\sP(X)]^{(L-1)} ~ \sP^{(ML)}(\err) \no \\  & = & \sP(\err) \left[
1 + o(1) \right] ~+~ [\sP(X)]^{(L-1)} ~ \sP^{(ML)}(\err) \label{exact} \\[0.1in]
& \le & e^{-N [E_1(R_1,T) + T]} \left[ 1 + o(1) \right] ~+~ e^{-N[ E_r(R_1) +
(L-1)E_1(R_1,T)]} ~, \label{avgPe} \enqa
where the inequality follows from (\ref{upper}) and the random coding upper bound
on the ML decoding error probability \cite{Gall}. Letting $E_1(R_1,T) \to 0$ and
maximizing $T$ as before, we get the following error exponent
\[ E_{MD}(R,L) ~=~ \limsup_{N \to \infty} ~ -\frac{\ln \sP_{MD}(\tE)}{\ov{\tau}}
~\ge~ \min\{ E_F(R), E_r(R) \} ~=~ E_r(R), \] since the feedback
exponent $E_F(R)$ is known to be greater than the random coding
exponent $E_r(R)$. Thus by setting $E_1(R_1,T) \to 0$, as
suggested by intuitive reasoning, we find that memoryless decoding
does not give any improvement over ML decoding without feedback.
However, we can get better performance by optimizing the
expression in (\ref{avgPe}) w.r.t $T$ without letting $E_1(R_1,T)
\to 0$. From (\ref{avgPe}), it is clear that the optimal value of
the threshold $T^{*}$ is the one that yields
\[ E_1(R_1,T^{*}) + T^{*} ~ = ~ E_r(R_1) + (L-1) E_1(R_1,T^{*}) \]
\beq \label{cond} \Rightarrow \quad T^{*} ~ = ~ E_r(R_1) + (L-2)
E_1(R_1,T^{*}) ~. \enq Using this optimal value of $T^{*}$ in
(\ref{E1eqh}) and solving for $E_1(R_1,T^{*})$, we get \beq
\label{E1max} E_1(R_1,T^{*}) ~ = ~ \max_{0 \le s \le \rho \le 1,
\pv} \left( \frac{E_o(s,\rho, \pv) - \rho R_1
-sE_r(R_1)}{1+s(L-2)}  \right) ~. \enq Since $E_F(R_1) >
E_r(R_1)$, we have $E_1(R_1,T^{*}) > 0$ and hence $R \to R_1$ as
$N \to \infty$. Thus the error exponent of memoryless decoding is
lower bounded by \beqa E_{MD}(R,L) & \ge &  E_2(R,T^{*}) ~ = ~
E_1(R,T^{*}) + T^{*} ~ = ~ E_r(R)
+ (L-1) E_1(R,T^{*}) \no \\[0.1in] & = & E_r(R) + (L-1) \left[ \max_{0 \le s
\le \rho \le 1, \pv} \left( \frac{ E_o(s,\rho,\pv) - \rho R -s E_r(R)}{1+
s(L-2)} \right) \right]. \label{mdlb} \enqa
Since $E_1(R,T^{*}) > 0$, it is clear that the optimal threshold $T^{*}$
satisfies $0 \le T^{*} < E_F(R)$ and thus the lower bound on $E_{MD}(R,L)$ in
(\ref{mdlb}) is smaller than the feedback exponent $E_F(R)$.

We now derive an upper bound on $E_{MD}(R,L)$ from (\ref{exact}) as follows.
\beqa \sP_{MD}(\tE) & = & \sP(\err) \left[1 + o(1) \right] ~+~
[\sP(X)]^{(L-1)} ~ \sP^{(ML)}(\err) \no \\[0.05in] & \ge & [\sP(X)]^{(L-1)} ~
\sP^{(ML)}(\err) \no \\[0.05in] & \ge & [\sP(X)]^{(L-1)} ~ \left( e^{-N
E_{sp}(R_1)} \right) ~,\label{mdub} \enqa
where the last inequality follows from the sphere-packing lower bound on
the ML decoding error probability \cite{Gall}. It is easy to see that
the probability of erasure $\sP(X)$ of the decoder in (\ref{Rm}) decreases
when the threshold parameter $T$ is decreased. Thus the probability of
erasure $\sP(X)|_{T=0}$ serves as a lower bound on $\sP(X)$ for any $T>0$.
In \cite{Viter}, upper and lower bounds on the erasure and error probabilities
are derived using a theorem of Shannon {\em et al} \cite{Shan67}. From
equations (10) and (11) in \cite{Viter}, we have
\[ \frac{1}{4M} \sum_{m=1}^M \exp \left[ \mu_m(s) - s\mu_m^{'}(s) - s\sqrt{2
\mu_m^{''}(s)} \right] ~<~ \sP(X) + \sP(\err) ~\le~ \frac{1}{M}
\sum_{m=1}^M \exp \left[ \mu_m(s) - s\mu_m^{'}(s) \right] ~,\] and
\beqn \frac{1}{4M} \sum_{m=1}^M \exp \left[ \mu_m(s) +(1-s)
\mu_m^{'}(s) - (1-s)\sqrt{2 \mu_m^{''}(s)} \right] &<& \sP(\err)
\\ & \le & \frac{1}{M} \sum_{m=1}^M \exp \left[ \mu_m(s) + (1-s)
\mu_m^{'}(s) \right] ~, \enqn where
\[ \mu_m(s) ~=~ \ln \int p(\ym|\xm)^{(1-s)} \left[ \sum_{m_1 \ne m}
p(\ym|{\mathbf x}_{m_1}) \right]^s \ud \ym ~. \]
It is clear from equation (8) in \cite{Viter} that the threshold parameter
$T$ is related to the parameter $\mu_m(s)$ by $\mu_m^{'}(s)=-NT$. Thus the
condition $T=0$ corresponds to the condition $\mu_m^{'}(s) = 0$. Moreover, it
is shown in \cite{Viter} that $\mu_m(s)$ and $\mu_m^{''}(s)$ are also proportional
to $N$. Using this fact and the condition $\mu_m^{'}(s) = 0$ in the above
expressions for the upper and lower bounds on $\sP(X)$ and $\sP(\err)$, we get
\beq \label{pxe}
\frac{1}{4M} \sum_{m=1}^M \exp \left[ \mu_m(s) \left( 1 + o\left( \frac{1}
{\sqrt{N}} \right) \right) \right] ~ < ~ \sP(X) + \sP(\err) ~\le~ \frac{1}{M}
\sum_{m=1}^M \exp \left[ \mu_m(s) \right] ~, \enq
and \beq \label{pe}
\frac{1}{4M} \sum_{m=1}^M \exp \left[ \mu_m(s) \left( 1 + o\left(
\frac{1}{\sqrt{N}} \right) \right) \right] ~<~ \sP(\err) ~\le~ \frac{1}{M}
\sum_{m=1}^M \exp \left[ \mu_m(s) \right] ~. \enq
It is clear from (\ref{pxe}) and (\ref{pe}) that when $T=0$, the exponents of
the upper and lower bounds coincide as $N \to \infty$, and more importantly,
the exponent of the erasure probability $\sP(X)$ is the same as that of the
error probability $\sP(\err)$. These exponents are further equal to the
exponent of the ML decoding error probability since $\sP(\err) \le \sP^{(ML)}
(\err) \le \sP(\err) + \sP(X)$. Using this fact and the sphere-packing lower
bound on the ML decoding error probability in (\ref{mdub}), we get
\[ \sP_{MD}(\tE) ~\ge ~ e^{-N L E_{sp}(R_1)} \quad \Rightarrow ~~ E_{MD}(R,L)
~\le ~ L E_{sp}(R) ~, \] since $R \to R_1$ as $N \to \infty$.
\end{proof}

From Theorem~\ref{md}, it is clear that ARQ with memoryless
decoding does not achieve Forney's error exponent $E_F(R)$ when
the maximum number of ARQ rounds $L$ is constrained, at least at
high rates for which $L E_{sp}(R) < E_F(R)$. As expected, when $L
\to \infty$, the lower bound on the error exponent in (\ref{errexp})
becomes
\[ \lim_{L \to \infty} E_{MD}(R,L) ~\ge~  \max_{0 \le s \le \rho \le 1, \pv} \left(
\frac{ E_o(s,\rho,\pv) - \rho R}{s} \right) ~ = ~ E_F(R). \]

\subsection{Incremental Redundancy ARQ}  \label{memL}
We now derive a lower bound on the error exponent of incremental
redundancy ARQ. In IR-ARQ, the transmitter, upon receiving a NACK
message, transmits $N$ {\bf new} coded symbols (derived from the
same message). Since our results hinge on random coding arguments,
these new symbols are obtained as i.i.d. realizations from the
channel capacity achieving distribution. The decoder does not
discard the received observations in the case of an erasure and
uses the received sequences of all the ARQ rounds jointly to
decode the transmitted message. The following erasure decoding
rule is employed by the receiver: After the $k^{th}$ ARQ round,
the decoder decides on codeword $\xm$ iff $\ym \in {\mathcal
R}^{'}_m$, where \beq \label{newdec} {\mathcal R}^{'}_m ~=~
\left\{ \ym ~:~ \frac{p(\ym|\xm)}{\sum_{i \ne m} p(\ym| {\mathbf
x}_i)} ~\ge ~e^{kN T_k}  \right\} ~, \enq and $\ym$, $\{ {\mathbf
x}_i \}$ are vectors of length $kN$, which contain the received
sequences and transmitted codewords (respectively) corresponding
to the $k$ ARQ rounds. If no codeword satisfies the above
condition, then an erasure is declared by the decoder. It is clear
that our formulation allows for varying the threshold $T_k$ as a
function of the number of ARQ rounds $k$. Using thresholds
$\{T_k\}$ that decrease with the number of ARQ rounds $k$ makes
intuitive sense since the probability of error will be dominated
by small values of $k$ (initial ARQ rounds), and hence, one needs
to use higher thresholds for these $k$ values to reduce the
overall probability of error. We let $E_k$ denote the event that
the decoder declares an erasure during {\em all} the first $k$ ARQ
rounds. We also let $E_0 = {\mathbf \phi}$ (the empty set). The
probability of erasure and error of the decoder in the $k^{th}$
ARQ round will thus be denoted by $\sP_{(k)}(X|E_{(k-1)})$ and
$\sP_{(k)}(\err|E_{(k-1)})$, respectively. Here the subscript
$(k)$ is used to highlight the fact that the decoder uses a
received sequence of length $kN$ for decoding in the $k^{th}$ ARQ
round. We are now ready to state our main result in this section.

\begin{theorem} \label{mem}
The error exponent $E_{IR}(R,L)$ achieved by IR-ARQ, under a deadline
constraint $L$, is given by\footnote{Replacing the random coding
exponent $E_r(R)$ by the expurgated exponent $E_{ex}(R)$ may yield
a tighter lower bound at low rates.} \beq \label{errgen} E_{IR}(R,L)
~ \ge ~ \min \left\{ E_F(R)~, ~ L E_r(R/L) \right\} ~, \quad 0 \le
R \le C. \enq
\end{theorem}
\begin{proof}
The average decoding delay for IR-ARQ is given by \beqa \ov{\tau}
& = & \sum_{k=1}^L kN ~ \Prob(\textrm{Transmission stops after
 $k$ ARQ rounds}) \no \\ & = & \sum_{k=1}^{L-1} kN \left( \prod_{i=1}^{k-1}
 \sP_{(i)}(X|E_{(i-1)}) \right) \left[ 1 - \sP_{(k)}(X|E_{(k-1)}) \right] \no \\
 & & \qquad + ~ LN \left( \prod_{i=1}^{L-1}  \sP_{(i)}(X|E_{(i-1)}) \right) \no \\
& = & \sum_{k=0}^{L-1} (k+1)N \left( \prod_{i=1}^{k}
\sP_{(i)}(X|E_{(i-1)}) \right) ~- ~\sum_{k=1}^{L-1} kN \left(
\prod_{i=1}^{k} \sP_{(i)}(X|E_{(i-1)}) \right) \no \\  & = & N
\left[ 1 + \sum_{k=1}^{L-1} \left( \prod_{i=1}^{k}
\sP_{(i)}(X|E_{(i-1)})\right) \right] \no \\ & \le & N \left[ 1 +
\sum_{k=1}^{L-1} \sP(X) \right] ~ \le ~ N \left[ 1 + L \sP(X)
\right] . \label{avgn} \enqa Since $\sP(X) \le e^{-N E_1(R_1,T)}$,
it follows that $\ov{\tau} \to N$ (and hence the average effective
transmission rate $R \to R_1$) as $N \to \infty$ when $E_1(R_1,T)
> 0$. The average probability of error of IR-ARQ is
given by \beqn \sP_{IR}(\tE) & = & \sum_{k=1}^{L} \Prob (\textrm{error
in the $k^{th}$ ARQ round}) \\ & = & \sum_{k=1}^{L-1}
\sP_{(k)}(\err,E_{(k-1)}) + \sP_{(L)}^{(ML)} (\err,E_{(L-1)}) \\ &
\le & \sum_{k=1}^{L-1} \sP_{(k)}(\err) ~+~ \sP_{(L)}^{
(ML)}(\err)~, \enqn where $\sP_{(k)}(\err)$ refers to the
probability of error when the decoder always waits for $kN$
received symbols before decoding. Following the derivation in
\cite{Forn}, it can easily be seen that for the thresholds
$\{T_k\}$ used in the decoding rule (\ref{newdec}), we have \beq
\label{ub} \sP_{(k)}(X) ~\le ~e^{-kN E_1(R_1/k,T_k)} \quad
\textrm{and} \quad \sP_{(k)}(\err) ~\le ~e^{-kN [ E_1(R_1/k,T_k) +
T_k]} ~. \enq Using this and the fact that $\sP_{(L)}^{(ML)}(\err)
\le e^{-LN E_r(R_1/L)}$, the average probability of error of
IR-ARQ can be upper bounded by \beq \label{Pegen} \sP_{IR}(\tE) ~ \le ~
\sum_{k=1}^{L-1} e^{-kN [ E_1(R_1/k,T_k) + T_k]} ~ + ~ e^{-LN
E_r(R_1/L)} ~. \enq Thus the error exponent achieved by IR-ARQ is
lower bounded by
\[ E_{IR}(R,L) ~ = ~ \limsup_{N \to \infty} ~-\frac{\ln \sP_{IR}(\tE)}{\ov{\tau}} ~ \ge ~
\min \left( L E_r(R/L), ~ \left\{ ~k[E_1(R/k,T_k) + T_k] ~\right\}_{k=1}^{L-1}
\right). \] Taking $T_k = (T/k)$, $\forall k \in \{1, \cdots, (L-1)\}$, we get
\beqn E_{IR}(R,L) & \ge & \min \left( L E_r(R/L), ~ \left\{ ~k E_1(R/k,T/k) + T ~
\right\}_{k=1}^{L-1} \right) \\[0.05in] & = & \min \left( L E_r(R/L),~ E_1(R,T) + T
\right) , \enqn
where the last equality follows from the fact that $E_1(R/k,T/k)$ is an
increasing function of $k$. Letting $E_1(R,T) \to 0$ and maximizing $T$, we get
\[ E_{IR}(R,L) ~\ge ~ \min \left( L E_r(R/L), E_F(R) \right) ~. \]
\end{proof} \\
From Theorem~\ref{mem}, it is clear that if the deadline
constraint $L$ is large enough to satisfy \beq \label{Lval} L
E_r(R/L) ~\ge~ E_F(R) ~, \enq then IR-ARQ achieves the feedback
exponent $E_F(R)$ at rate $R$. In the following section, we
quantify the gains achieved by IR-ARQ, as compared with memoryless
decoding, for specific channels.

\section{Examples}  \label{examples}

\subsection{The Binary Symmetric Channel (BSC)}
Here, we compare the error exponents achievable by memoryless
decoding and IR-ARQ over a BSC with crossover probability
$\epsilon$. The bounds on the error exponents in (\ref{errexp})
and (\ref{errgen}) are plotted for a BSC with $\epsilon = 0.15$ in
Figs.~\ref{d2comp_bsc} and \ref{d4comp_bsc} for $L=2$ and $L=4$,
respectively. The ML decoding error exponent (corresponding to the
case $L=1$) and the feedback exponent $E_F(R)$ are also plotted
for comparison purposes. From Fig.~\ref{d2comp_bsc}, we find that
when $L=2$, memoryless decoding achieves an error exponent that is
strictly sub-optimal to the feedback exponent $E_F(R)$ for all $R
\ge 0.006$. On the other hand, IR-ARQ achieves $E_F(R)$ for $0.18
\le R \le C$. Moreover, it performs strictly better than
memoryless decoding for all $R \ge 0.057$. When $L=4$, from
Fig.~\ref{d4comp_bsc}, we find that the error exponent for the
memoryless decoder is strictly sub-optimal, as compared with
$E_F(R)$, for $R \ge 0.141$, while IR-ARQ achieves $E_F(R)$ for all
rates below capacity. Finally, we note that even when $L=100$,
memoryless decoding is still strictly sub-optimal, as compared
with IR-ARQ, for all rates $0.38 \le R \le C=0.39$.

Now, we elaborate on our observation from Fig.~\ref{d4comp_bsc}
that $L=4$ is sufficient to achieve $E_F(R)$ with IR-ARQ when
$\epsilon=0.15$. In particular, we wish to investigate the
existence of a finite value for $L$ such that $E_F(R)$ is achieved
by IR-ARQ universally (i.e., for all $0\leq \epsilon\leq 0.5$ and
all rates below capacity). Towards this end, we derive an upper
bound on the {\em minimum} required deadline constraint $L_{req}$
for a given BSC($\epsilon$). From (\ref{errgen}), it is clear that
$L_{req}$ is upper bounded by the minimum value of $L$ required to
satisfy $L E_r(R/L) \ge E_F(R)$ for all $0 \le R \le C$. We first
prove the following result.

\begin{lemma} \label{Lub}
A sufficient condition for ensuring that $L E_r(R/L) \ge E_F(R)$ for all rates
$0 \le R \le C$ for a BSC is given by $L E_r(0) \ge E_F(0)$.
\end{lemma}
\begin{proof}
It has been shown in \cite{Gall} that both the random coding exponent $E_r(R)$
and the feedback exponent $E_F(R)$ are decreasing functions of $R$. Since
\beq \label{maxrho}
L E_r(R/L) ~= ~\max_{0 \le \rho \le 1} \left\{ L E_o(\rho) - \rho R \right\} ~,
\enq
its slope at a given rate $R$ is given by (following the steps in equations
(5.6.28--5.6.33) in \cite{Gall})
\[ \frac{\partial \left( L E_r(R/L) \right)}{\partial R} ~= ~- \rho^{*}(R) ~
\ge ~ - 1 ~,\]
where $\rho^{*}(R)$ is the value of $\rho$ that maximizes the RHS of
(\ref{maxrho}) for rate $R$. For a BSC, it is shown in \cite{Forn} that
the feedback exponent can be expressed as
\beq \label{maxrhof}
E_F(R) ~= ~(C - R) ~+~ \max_{\rho \ge 0} \{ E_o(\rho) - \rho R \} ~. \enq
Hence the slope of $E_F(R)$ at a given rate $R$ is given by
\[ \frac{\partial E_F(R)}{\partial R} ~= ~- \left(1 + \rho^{'}(R) \right) ~
\le ~ -1 ~, \]
where $\rho^{'}(R)$ is the value of $\rho$ that maximizes the RHS of
(\ref{maxrhof}) for rate $R$.
Hence it is clear that for any value of $R$, the rate of decrease of the
feedback exponent $E_F(R)$ is higher than that of $L E_r(R/L)$. It is shown
in \cite{Forn} that $E_F(C) = E_r(C) = 0$. Since $E_r(R)$ is a decreasing
function of $R$, we know that $E_r(C/L) > E_r(C) = 0$. Thus, when $R=C$,
we have $L E_r(C/L) > E_F(C)$.  Now, if the value of $L$ is chosen such that
$L E_r(0) > E_F(0)$, it is clear that the curve $L E_r(R/L)$ lies strictly
above the curve $E_F(R)$ in the range $0 \le R \le C$. This directly follows
from the fact that the feedback exponent $E_F(R)$ decreases faster than
$L E_r(R/L)$. Hence the condition $L E_r(0) \ge E_F(0)$ is sufficient to
guarantee that $L E_r(R/L) \ge E_F(R)$ for all $0 \le R \le C$.
\end{proof}

The above lemma shows that for any BSC($\epsilon$), an upper bound
on $L_{req}$ depends only on the values of $E_F(R)$ and $E_r(R)$
at $R=0$. From the results in \cite{Gall}, it can be shown that
\beq \label{exp0} E_r(0) ~=~ \ln 2 - \ln \left( 1 + 2
\sqrt{\epsilon (1-\epsilon)} \right) \quad \textrm{and} \quad
E_F(0) ~ = ~ C - \ln 2 - \ln \left( \sqrt{\epsilon (1-\epsilon)}
\right) ~. \enq Using Lemma~\ref{Lub} and (\ref{exp0}), we find
that a deadline constraint of $L=4$ is enough to achieve the
feedback exponent $E_F(R)$ at all rates below capacity for {\em
any} BSC with crossover probability $0.05 \le \epsilon \le 0.5$.
However, the upper bound on $L_{req}$, derived using
Lemma~\ref{Lub}, becomes loose as $\epsilon \to 0$. To overcome
this limitation, we use the expurgated exponent $E_{ex}(R)$
\cite{Gall} instead of the random coding exponent $E_r(R)$ at low
rates. Using numerical results, we find that the actual value of
the minimum required deadline constraint is $L_{req}=3$ for all
BSCs with $\epsilon \le 0.025$, and $L_{req}=4$ otherwise.

\subsection{The Very Noisy Channel (VNC)}
As noted in \cite{Gall}, a channel is very noisy when the probability of
receiving a given output is almost independent of the input, i.e., when
the transition probabilities of the channel are given by
\[ p_{jk} ~ = ~ \omega_j \left( 1 + \epsilon_{jk} \right) ~,\]
where $\{ \omega_j \}$ denotes the output probability
distribution, and $\{ \epsilon_{jk} \}$ are such that
$|\epsilon_{jk}| \ll 1$ for all $j$ and $k$, and $\sum_{j}
\omega_j \epsilon_{jk} = 0$, $\forall k$. We plot the bounds on
the error exponents given in (\ref{errexp}) and (\ref{errgen}),
derived from the results in \cite{Forn}, in Figs.~\ref{d2comp_vnc}
and \ref{d4comp_vnc} for a VNC with capacity $C=1$ for $L=2$ and
$L=4$ respectively. From the plots, it is clear that memoryless
decoding is strictly sub-optimal to IR-ARQ for all rates $R \ge
0.12$ (with $L=2$) and $R \ge 0.25$ (with $L=4$). Moreover, it is
evident that $L=4$ is sufficient for IR-ARQ to achieve the
feedback exponent $E_F(R)$ for all rates below capacity. This
observation motivates the following result.

\begin{lemma}
For the very noisy channel, a deadline constraint of $L=4$ is enough for
the proposed incremental redundancy scheme to achieve the feedback
exponent $E_F(R)$ for all rates $0 \le R \le C$.
\end{lemma}
\begin{proof}
For a VNC, the random coding exponent is given by \cite{Forn}
\beq \label{vncEr}
E_r(R) ~ = ~ \left\{ \begin{array}{cl} \left(\frac{C}{2} - R \right) ~,
& 0 \le R \le \frac{C}{4} \\[0.05in] (\sqrt{C} - \sqrt{R})^2 ~, & 
\frac{C}{4} \le R \le C \end{array} \right. . \enq
Thus, under the deadline constraint $L=4$, we have
\[ 4 E_r(R/4) ~ = ~ 4 \left(\frac{C}{2} - \frac{R}{4}\right) ~=~ 2C - R
~, \qquad 0 \le R \le C. \] Also \[ E_F(R) ~=~ (C - R) + (\sqrt{C} -
\sqrt{R})^2 ~\le~ (C-R)+ (\sqrt{C})^2 ~=~ 4 E_r(R/4) ~. \] Putting
$L=4$ in (\ref{errgen}), the error exponent of IR-ARQ is given by
\beq \label{vncmem} E_{IR}(R,4) ~\ge~ \min \left\{ E_F(R)~, ~ 4
E_r(R/4) \right\} ~=~ E_F(R) ~. \enq Thus, for a VNC, it is clear
that a deadline constraint of $L=4$ is enough for IR-ARQ to
achieve the feedback exponent $E_F(R)$ at all rates below
capacity.
\end{proof}

\subsection{The Additive White Gaussian Noise (AWGN) channel}

The random coding and expurgated exponents for an AWGN channel with a
Gaussian input of power $A$ and unit noise variance, are given in
\cite{Gall}. The sphere-packing exponent of the AWGN channel is derived
in \cite{Shan59,Foss,Divs}. The parameter $E_o(s,\rho,\pv)$ in the lower
bound in (\ref{errexp}) is replaced by $E_o(s,\rho,t)$ which, following
the steps in the derivation of the random coding exponent in \cite{Gall},
is given by
\beqn E_o(s,\rho,t) ~=~ (1+\rho) t A + \left(\frac{1}{2}\right) \log (1-2tA)
+ \left(\frac{\rho}{2}\right) \log \left( 1- 2tA+ \frac{sA}{\rho}
\right) \\ \qquad + ~\left(\frac{1}{2}\right) \log \left( 1 + \frac{sA \left(
1-s-\frac{s}{\rho} \right)}{1 - 2tA + \frac{sA}{\rho}} \right). \enqn
The feedback exponent for the AWGN channel is then given by \cite{Forn,Gall}
\[ E_F(R) = \max_{0 \le s \le \rho \le 1, t \ge 0} \left( 
\frac{E_o(s,\rho,t)- \rho R}{s} \right) ~. \]
We plot the bounds on the error exponents, given in (\ref{errexp}) and
(\ref{errgen}), in Figs.~\ref{d2comp} and \ref{d4comp} for an AWGN channel
with signal-to-noise ratio $A=3$ dB for the deadline constraints $L=2$
and $L=4$ respectively. The plots clearly indicate that memoryless decoding
is strictly sub-optimal to IR-ARQ for all rates $R \ge 0.19$ (with $L=2$)
and $R \ge 0.46$ (with $L=4$). Moreover, when $L=4$, the proposed IR-ARQ
scheme achieves the feedback exponent $E_F(R)$ for all rates below
capacity (at the moment, we do not have a proof that this observation
holds universally as in the case of BSCs).

\section{Conclusions}  \label{conc}
We considered the error exponents of memoryless ARQ channels with
an upper bound $L$ on the maximum number of re-transmission
rounds. In this setup, we have established the superiority of
IR-ARQ, as compared with Forney's memoryless decoding. For the BSC
and VNC, our results show that choosing $L=4$ is sufficient to
ensure the achievability of Forney's feedback exponent, which is
typically achievable with memoryless decoding in the asymptotic
limit of large $L$. Finally, in the AWGN channel, numerical
results also show the superiority of IR-ARQ over memoryless
decoding, in terms of the achievable error exponent.

\newpage
\bibliography{arq_exponent}
\bibliographystyle{ieeetr}

\newpage
\renewcommand{\thefigure}{\arabic{figure}(\alph{subfig})}
\setcounter{subfig}{1}
\begin{figure}
\centering
\includegraphics[height=0.45\textheight]{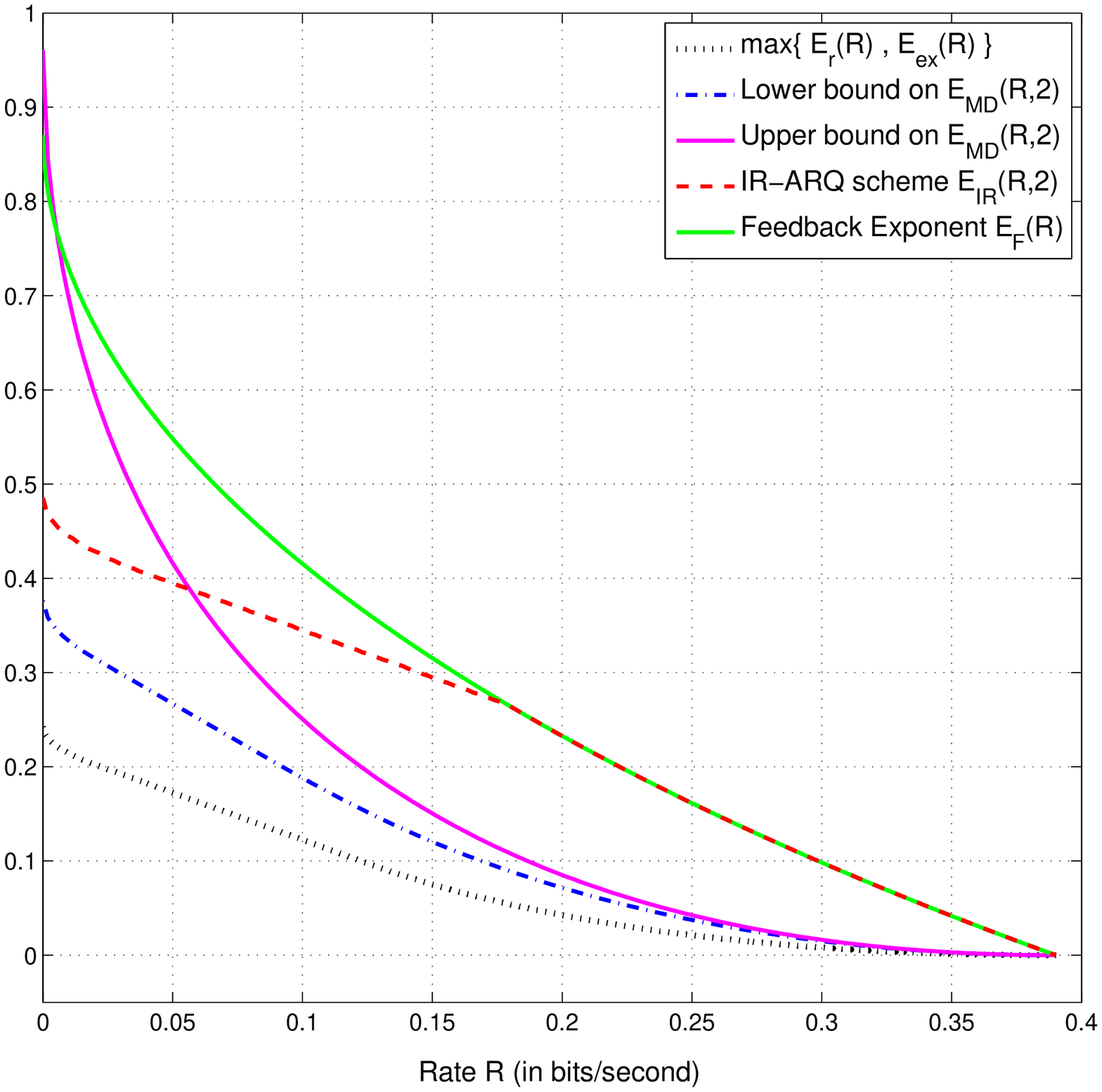}
\caption{Comparison of the error exponents for a BSC with $\epsilon=0.15$ and $L=2$
\label{d2comp_bsc}}
\end{figure}

\addtocounter{figure}{-1}
\addtocounter{subfig}{1}
\begin{figure}
\centering
\includegraphics[height=0.45\textheight]{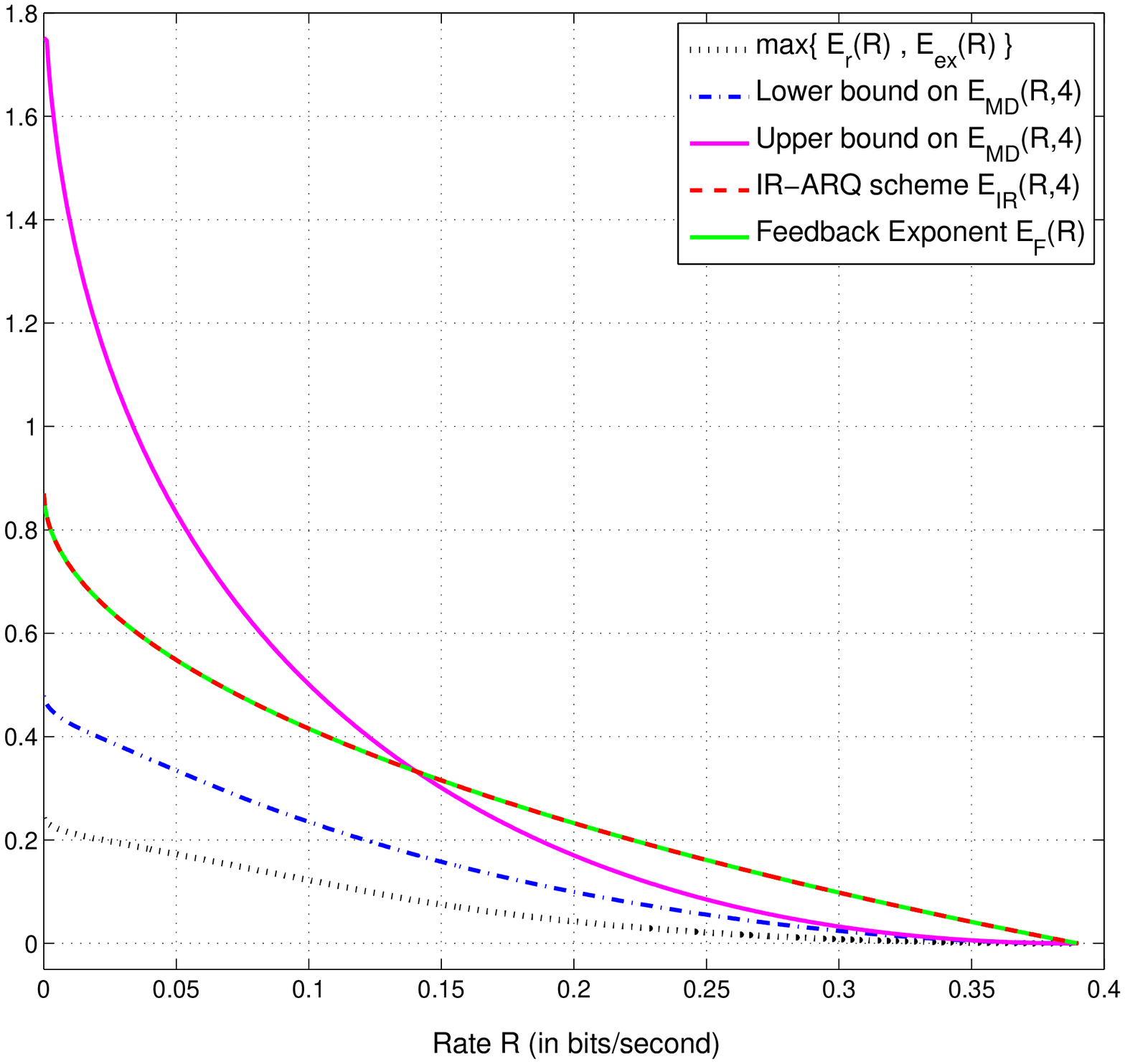}
\caption{Comparison of the error exponents for a BSC with $\epsilon=0.15$ and $L=4$
\label{d4comp_bsc}}
\end{figure}

\setcounter{subfig}{1}
\begin{figure}
\centering
\includegraphics[height=0.45\textheight]{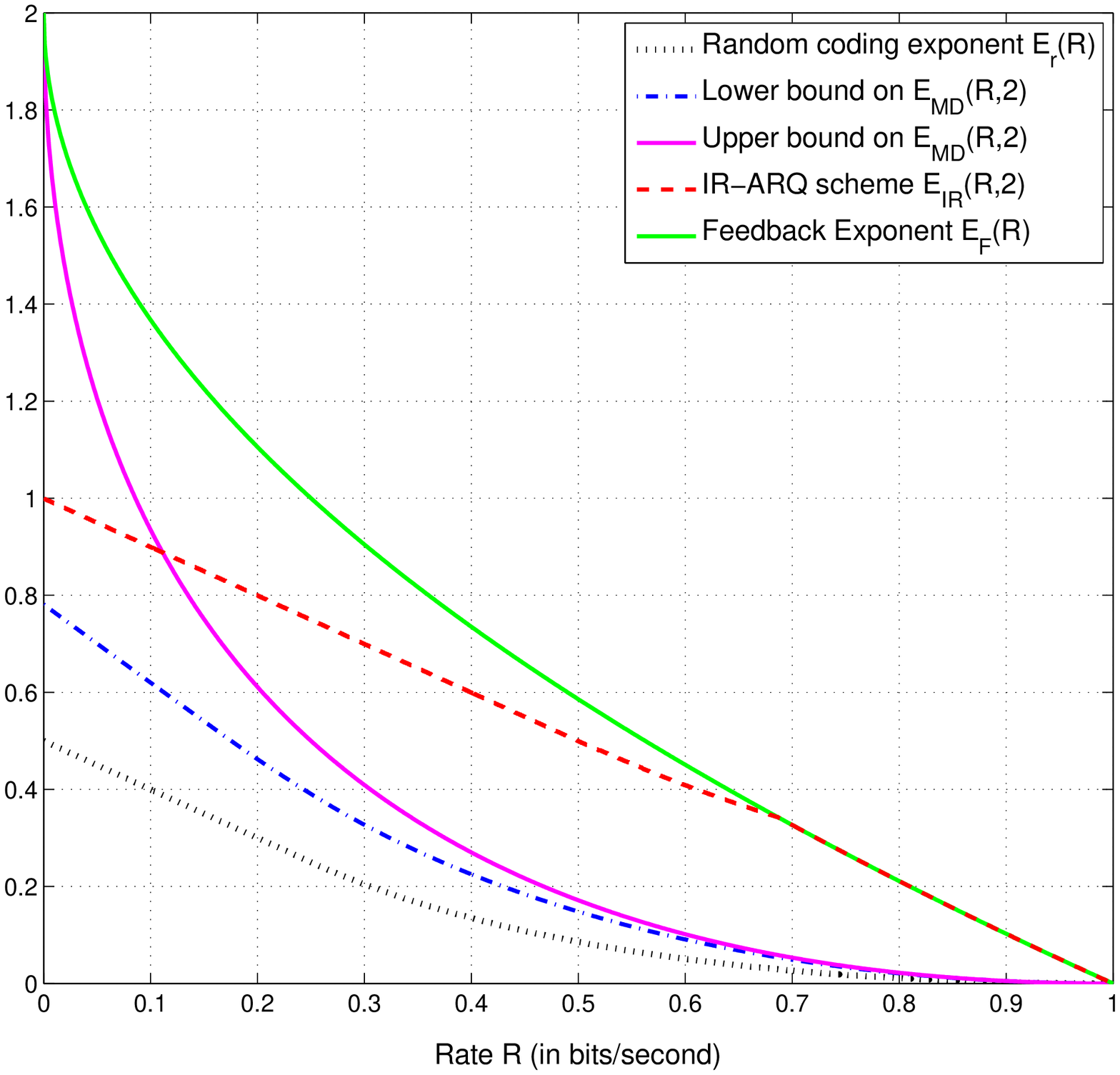}
\caption{Comparison of the error exponents for a VNC with $C=1$ and $L=2$
\label{d2comp_vnc}}
\end{figure}

\addtocounter{figure}{-1}
\addtocounter{subfig}{1}
\begin{figure}
\centering
\includegraphics[height=0.45\textheight]{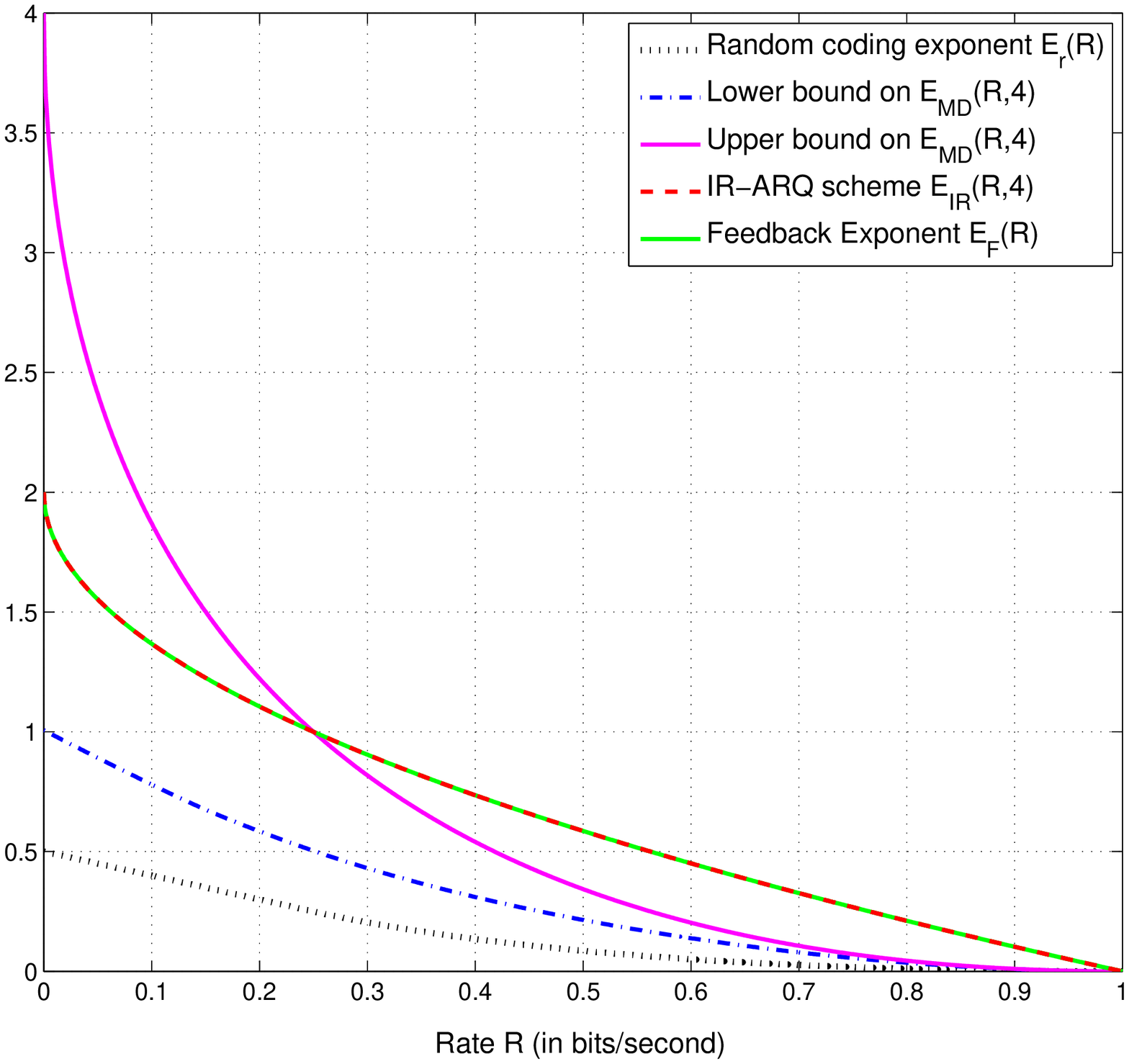}
\caption{Comparison of the error exponents for a VNC with $C=1$ and $L=4$
\label{d4comp_vnc}}
\end{figure}

\setcounter{subfig}{1}
\begin{figure}
\centering
\includegraphics[height=0.45\textheight]{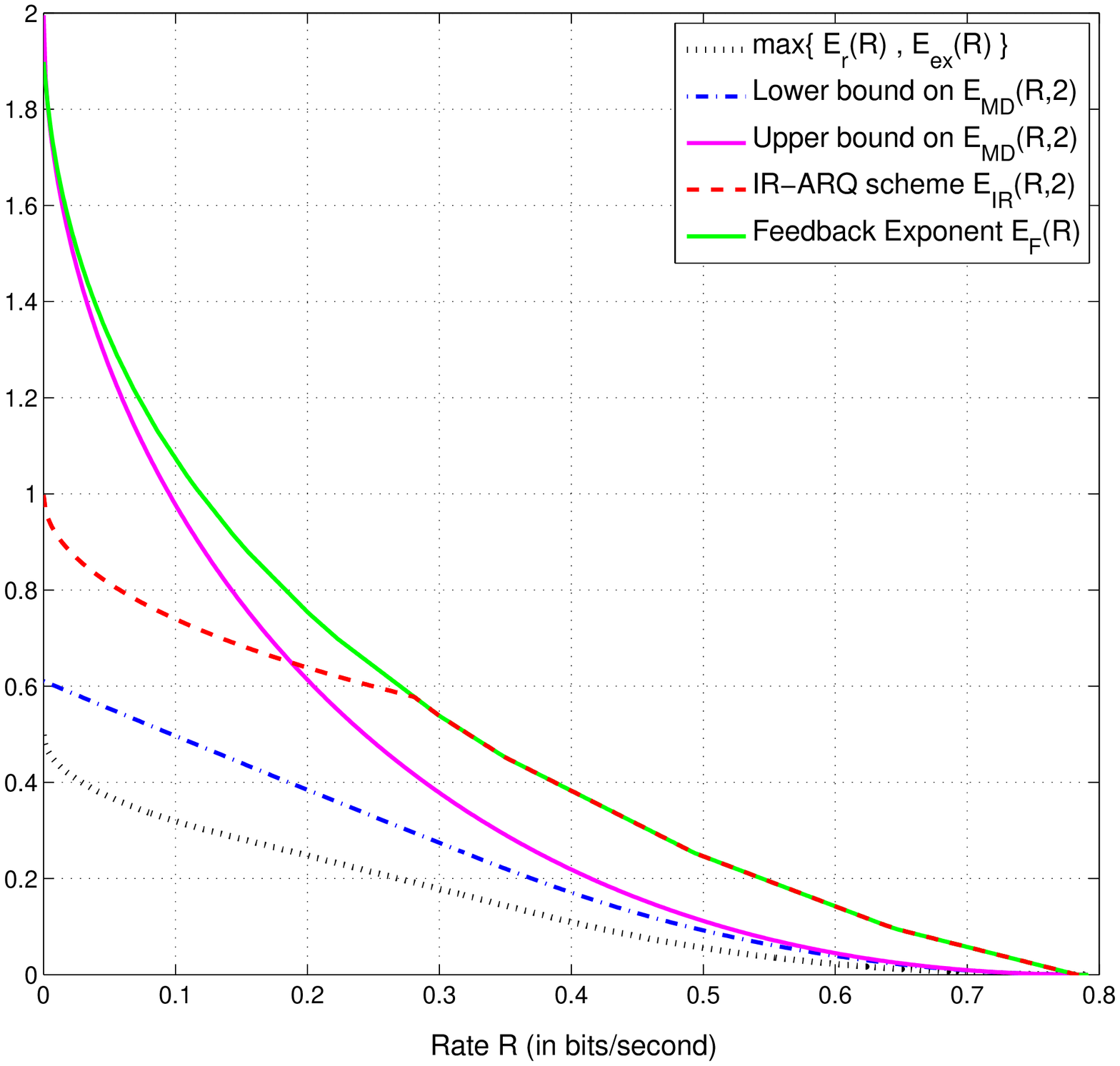}
\caption{Comparison of error exponents for an AWGN channel with SNR~=~3 dB and $L=2$
\label{d2comp}}
\end{figure}

\addtocounter{figure}{-1}
\addtocounter{subfig}{1}
\begin{figure}
\centering
\includegraphics[height=0.45\textheight]{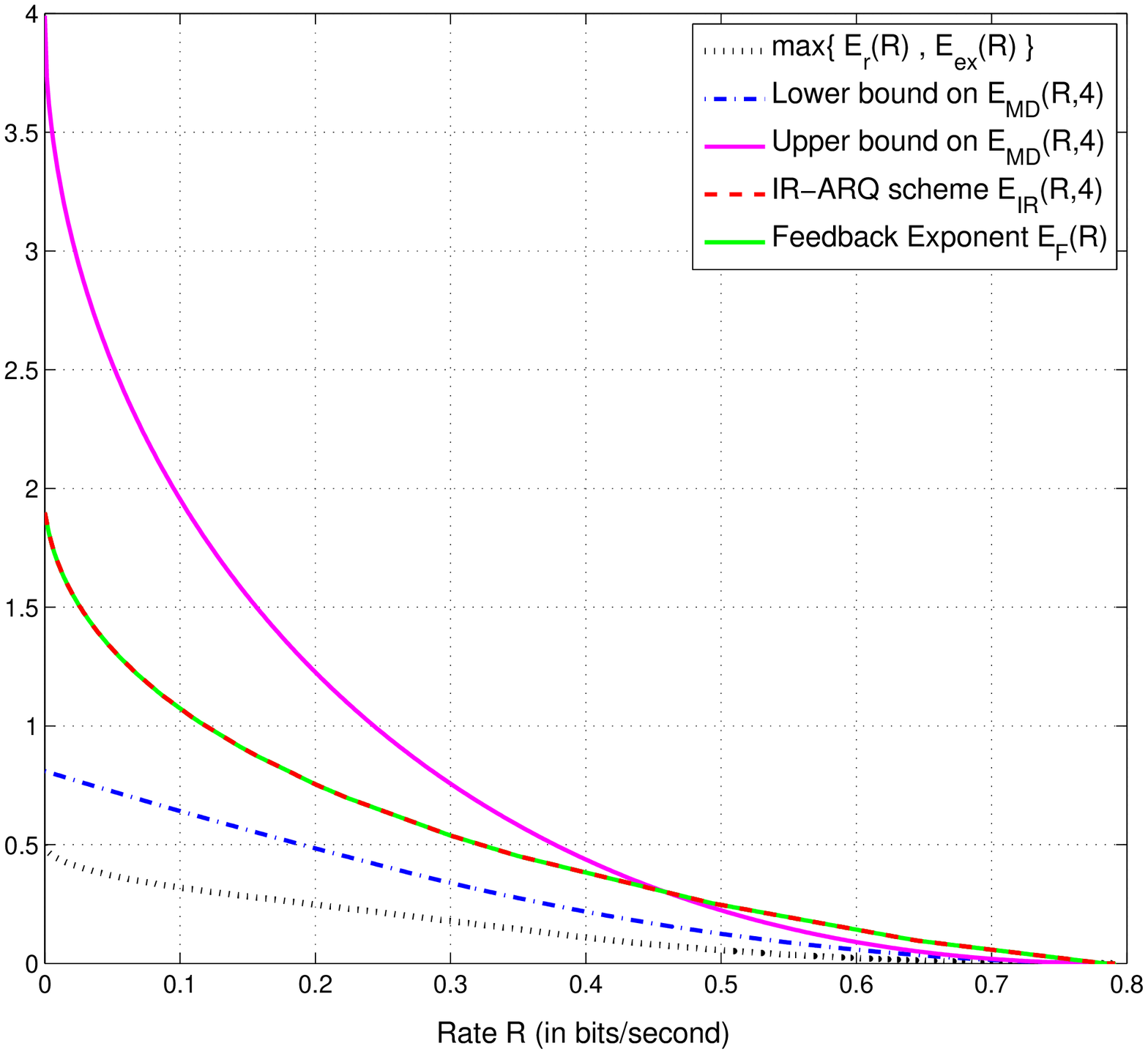}
\caption{Comparison of error exponents for an AWGN channel with SNR~=~3 dB and $L=4$
\label{d4comp}}
\end{figure}

\end{document}